\documentclass[12pt]{article}
\usepackage{amsfonts,amssymb,epsfig}
\usepackage[latin1]{inputenc}
\usepackage[english]{babel}
\usepackage{color}
\newtheorem{thm}{Th\'eor\`eme}[section]
\newtheorem{cor}[thm]{Corollaire}
\newtheorem{lem}[thm]{Lemme}
\newtheorem{pro}[thm]{Proposition}
\newtheorem{dfn}[thm]{D\'efinition}
\newtheorem{rmq}[thm]{Remark}
\newtheorem{expl}[thm]{Exemple}

\def\dessous#1\sous#2{\mathrel{\mathop{\kern0pt#2}\limits_{#1}}}

\newcommand{\R}{\mathbb R}
\newcommand{\C}{\mathbb C}

\newcommand{\1}{1 \! \! {\rm I}}

\newcommand{\beq}{\begin{eqnarray}}
\newcommand{\eeq}{\end{eqnarray}}
\newcommand{\bpro}{\begin{pro}}
\newcommand{\epro}{\end{pro}}
\newcommand{\blem}{\begin{lem}}
\newcommand{\elem}{\end{lem}}
\newcommand{\bdfn}{\begin{dfn}}
\newcommand{\edfn}{\end{dfn}}
\newcommand{\bcor}{\begin{cor}}
\newcommand{\ecor}{\end{cor}}
\newcommand{\bthm}{\begin{thm}}
\newcommand{\ethm}{\end{thm}}
\newcommand{\bex}{\begin{expl}}
\newcommand{\eex}{\end{expl}}
\newcommand{\brmq}{\begin{rmq}}
\newcommand{\ermq}{\end{rmq}}
\newcommand{\benum}{\begin{enumerate}}
\newcommand{\eenum}{\end{enumerate}}
\newcommand{\bitem}{\begin{itemize}}
\newcommand{\eitem}{\end{itemize}}
\begin{document}

\begin{titlepage}
   \begin{center}
     {\ }\vspace{0.5cm}
       {\bf {Quaternionic  vector coherent states for
 spin-orbit interactions}}

 \vspace{1.5cm}

\vspace{1.5cm}

Isiaka Aremua$^{ a}$ and Mahouton Norbert Hounkonnou$^\dag$

\vspace{1cm}

     \vspace{1cm}

       {\em  International Chair of Mathematical Physics
and Applications} \\
{\em ICMPA-UNESCO Chair}\\
{\em University of Abomey-Calavi}\\
{\em 072 B.P. 50 Cotonou, Republic of Benin}\\
{\em E-mail: {\tt $^\dag$norbert.hounkonnou@cipma.uac.bj\footnote{Corresponding author
 (with copy to hounkonnou@yahoo.fr)}, $^a$ claudisak@yahoo.fr,
}}

     \vspace{2.0cm}
           \today
          \begin{abstract}
\noindent
    This work addresses  the study of two-dimensional electron gas in a perpendicular magnetic field
in the presence of both Rashba and Dresselhaus  spin-orbit (SO) interactions,  with  effective Zeeman coupling. Exact analytical expressions are found 
for the
 eigenvalue problem giving
the Landau levels and the associated eigenstates, thanks to an appropriate method of parametrization used to
quantize the physical system. Such a system exhibits interesting properties of quaternions
for which vector coherent states are built. Similarity with the Weyl-Heisenberg group and some relevant 
properties of these vectors are discussed.
\end{abstract}

\end{center}

\end{titlepage}

\setcounter{footnote}{0}

\section{Introduction}
Spin orbit Hamiltonian systems  continue to attract increasing interest in recent years
 due to their prevalence in various physical applications as in
semiconductors and spintronics.  For  details in relevant
applications, see
 \cite{zarea} and references therein. To cite here a few, let us mention that
 in the paradigmatic Datta-Das spin transistor,
 the spin of the electron passing through
the device is controlled by the Rashba spin-orbit (SO)
interaction \cite{rashba}, which in turn can be varied by the application
of gate voltages. The Rashba interaction stems from
the structural inversion asymmetry (SIA) introduced by
a heterojunction or by surface or external fields. In semiconductors
with narrower energy gap (InGaAs, AlGaAs),
this effect is expected to be stronger. It has been shown
experimentally that the Rashba spin-orbit interaction
can be modified up to $50\%$
 by external gate voltages \cite{miller}, \cite{nitta}.
In addition to the Rashba coupling, there is also a
material-intrinsic Dresselhaus spin-orbit interaction. This
originates from the bulk inversion asymmetry (BIA) of the crystal,
and can be relatively large in semiconductors like InSb/InAlSb
\cite{dresselhaus}. Both Rashba and Dresselhaus interactions
contribute to the spin-dependent splitting of the band structure
of the host material, leading to dramatic spin-dependent phenomena
for electrons or holes in semiconductors.

Globally, the determination of the eigenvalues and eigenstates of
such systems is crucial for the  understanding and the computation
of a number of their important physical properties {\cite{shen}}.
 All these aspects motivate  the investigations
 of the properties of the SO Hamiltonian systems.
   Recently, standard and nonlinear vector coherent states (VCS)
have been constructed for  relevant generalized models in quantum optics
 for nonlinear spin-Hall effect \cite{joben1}
  and for the Jaynes-Cummings model in the rotating wave approximation \cite{joben2}. See  also references quoted
in
these works.
Despite all these investigations, it is worthy of attention to note that
only a few exactly analytically solvable eigenvalue problems associated to SO interactions exist in the literature.
Besides, in the most interesting case of the description of
the more general case, in which both Rashba and Dresselhaus
interactions are present with arbitrary strength, in the effective Zeeman coupling,  exact solutions to the eigenvalue problem are still lacking
to our best knowledge of the existing works. The present paper  aims, among others, at fulfilling such a gap.
 We give explicit exact analytical solutions
to the addressed eigenvalue problem and build quaternionic VCS 
(QVCS)  for the spinor Landau levels.
 Similarity properties in connection with the Weyl-Heisenberg group is examined.

The paper is organized as follows. Section $2$ deals with the model description in two
gauge symmetries corresponding to the two orientations of the magnetic field.
Then, the eigenvalue problem is addressed and solved. In section $3$, VCS and QVCS  
are formulated and some   physical     properties of the latter states  are investigated. Then, follows the conclusion.

\section{Physical Hamiltonians}

\subsection{Hamiltonian with Zeeman coupling}

The Hamiltonian  of a particle system  with mass $M$ and charge $e$ with Zeeman coupling $g$, in an
electromagnetic field can be defined
by 
 \beq{\label{h0}}
\mathcal H = \frac{1}{2M}\left({\bf p} - \frac{e}{c}  {\bf A({\bf
r},t)}\right)^{2}  - g\frac{\mu}{2}\sigma
\cdot \, {\bf B}
\eeq
where in the term $g \mu/2  \sigma . {\bf B} $, the symbol $\mu =
e \hbar / 2M$ denotes the Bohr magneton, $g$ the factor of
coupling  related to the particle   and $ \sigma = (\sigma_{x},
\sigma_{y}, \sigma_{z}) \equiv (\sigma_{1}, \sigma_{2},
\sigma_{3})$ the vector of Pauli spin matrices.

Let us first consider  the gauge symmetry 
${\bf A} = \left(-\frac{B}{2}y, \frac{B}{2}x \right)$. Then,
the generalized Hamiltonian (\ref{h0})  becomes for an electron with charge $-e$, 
\beq{\label{spo2}}
\mathcal H_{1_{0}}({\bf p}, {\bf r})  \equiv \mathcal H_{1_{0}} = \frac{1}{2M}\left[\left(p_{x} - \frac{eB}{2c}y\right)^{2}
+  \left(p_{y} + \frac{eB}{2c}x\right)^{2} \right]  - \frac{g\mu B}{2}\sigma_{z}.
\eeq

Following the general scheme of
canonical quantization,  we can  introduce the
coordinate and momentum operators $R_{i},\,P_{i},\, i=1,2$ corresponding to
 the classical coordinates $x,y$ and their  conjugate momenta $p_{x}, p_{y}$, respectively, so that
   the canonical commutation relations
\beq [R_{i},P_{j}] = \imath \hbar \delta_{ij}, \quad
{\mbox{for}}\quad R_{i} = X, Y \quad \mbox{and}\quad  P_{i} =
P_{x}, P_{y}, \; i = 1,2 \eeq
 hold.
Using the change of variables \beq{\label{ex1}} Z = X + \imath Y,
\qquad P_{z} = \frac{1}{2}(P_{x} - \imath P_{y}), \eeq
 and the  creation and annihilation  operators $b^{\dag}, b$  defined as
 \beq{\label{es2}}
b^{\dag} = -2\imath P_{\bar{z}} + \frac{eB}{2c}Z 
, \qquad b = 2\imath P_{z} + \frac{eB}{2c}\bar{Z},  
\eeq
   and such that
 \beq{\label{es04}}
[b, b^{\dag}] &=& 2M \hbar \omega_{c}
\eeq where 

 $\omega_{c} =
\frac{eB}{Mc}$ (the cyclotron frequency), the Hamiltonian
(\ref{spo2})  can be reexpressed in the manner 
 
\beq{\label{spo3}}
H_{1_{0}} = \hbar \omega_{c} \left(b'^{\dag}b' +
\frac{\1}{2}\right)\sigma_{0}
 + \hbar \omega_{c} \xi \sigma_{z}
, \quad \xi = -g\mu Mc/2 \hbar e,
\eeq

with $\sigma_{0} = \mathbb I_{2}$ the identity matrix, where the
 annihilation and creation operators $b$ and $b^{\dag}$ have been
simplified as follows:
 \beq{\label{es11}}
b = \sqrt{2M\hbar \omega_{c}}b', \qquad  b^{\dag} = \sqrt{2M\hbar
\omega_{c}}b'^{\dag}, \eeq
 with
\beq [b',b'^{\dag}] = \1.
\eeq

The eigenvalues related to $ H_{1_{0}}$ are given by
\beq \mathcal E^{\pm}_{n} = \hbar \omega_{c}\left(n + \frac{1}{2}\right) \mp \hbar \omega_{c} \xi.
\eeq
The related eigenstates, denoted by
$
\left(
\begin{array}{c}
|\Phi_{n}\rangle \\
0
\end{array}
\right),\,
\left(
\begin{array}{c}
0 \\
|\Phi_{n}\rangle
\end{array}
\right)$,
span the Hilbert space $\hat \mathfrak H := \C^{2} \otimes \mathfrak H$, with $\mathfrak
H:=span\{|\Phi_{n}\rangle\}^{\infty}_{n=0}$, $|\Phi_{n}\rangle$ corresponding to the number operator $N = b'^{\dag}b'$ orthonormalized eigenstates 
$|n\rangle = (1/\sqrt{n!})(b'^{\dag})^{n}|0\rangle$,  and  
\beq
\hat \mathfrak H =\left\{|\chi^{+} \otimes \Phi_{n} \rangle =  \left(
\begin{array}{c}
|\Phi_{n}\rangle \\
0
\end{array}
\right), \; |\chi^{-} \otimes \Phi_{n} \rangle =  \left(
\begin{array}{c}
0 \\
|\Phi_{n}\rangle
\end{array}
\right), \; n=0,1,2\dots \right\},
\eeq

where  { $\chi^{+}$ and  $\chi^{-}$, given by $ \chi^{+}= \left(
\begin{array}{c}
1 \\
0
\end{array}
\right),\, \chi^{-}= \left(
\begin{array}{c}
0 \\
1
\end{array}
\right)$, form an orthonormal basis of $\C^{2}$.

In opposite, in the gauge symmetry
$
{\bf A} = \left(\frac{B}{2}y,
-\frac{B}{2}x \right)$,  the Hamiltonian  is given by 
 \beq{\label{spo4}}
\mathcal H_{2_{0}}({\bf p}, {\bf r})  \equiv \mathcal H_{2_{0}} = \frac{1}{2M}\left[\left(p_{x}
 + \frac{eB}{2c}y\right)^{2} +  \left(p_{y} - \frac{eB}{2c}x\right)^{2} \right]  + \frac{g\mu
B}{2}\sigma_{z},
\eeq

with the same change of variables as in (\ref{ex1}).  The   creation and annihilation operators given, respectively, by
 \beq{\label{ei20}}
\hat d^{\dag} = 2\imath P_{z} - \frac{eB}{2c}\bar{Z} 
, \qquad \hat d = -2\imath P_{\bar{z}} - \frac{eB}{2c}Z 
\eeq

satisfy the following commutation relations
 \beq{\label{es4}}
[\hat d, \hat d^{\dag}] &=& 2 M \hbar \omega_{c}.
\eeq
The corresponding canonically quantized Hamiltonian can be
written as  
 \beq{\label{spo5}}
H_{2_{0}} = \hbar \omega_{c} \left(\hat d^{\dag}\hat d +
\frac{\1}{2}\right)\sigma_{0}
 + \hbar \omega_{c} \xi' \sigma_{z},
\eeq

 where $\xi' = -\xi$, through  the annihilation and creation operators $\hat d$ and
$\hat d^{\dag}$ built as  follows: \beq{\label{boson1}} \hat d =
\sqrt{2M\hbar \omega_{c}}\hat d', \qquad  \hat d^{\dag} = \sqrt{2M\hbar
\omega_{c}}\hat d'^{\dag}
\eeq
with
 \beq
[\hat d',\hat d'^{\dag}] =\1. \eeq 
The eigenvalues and eigenstates related to the Hamiltonian $H_{2_{0}}$ are obtained as previously.

\subsection{Eigenvalues and eigenstates of  general spin-orbit interactions  Hamiltonians}

Through a reparametrization using the Fock operators or by making use of  projections  and defining the Pauli matrices
 $\sigma_{\pm} = (\sigma_{x}\pm \imath \sigma_{y})/2$,
the generalization of Rashba $H_{R}$ and Dresselhaus $H_{D}$ SO interactions can be introduced
as:
\beq{\label{spi02}}
H_{R} = \imath \hbar \omega_{c} \gamma \left(\lambda(N){(b'^{\dag})}^{k}\sigma_{-} - {b'}^{k}\overline{\lambda(N)}\sigma_{+}\right),  
H_{D} = \hbar \omega_{c} \beta \left(b'^{k}\lambda(N)\sigma_{-} + \overline{\lambda(N)}{(b'^{\dag})}^{k} \sigma_{+}\right)
\eeq
such that,  for two values of the parameter $\varepsilon$, there follow
\bitem
\item [(i)] $\varepsilon = +,$

\beq{\label{spi03}}
B^{+}_{k,+} = b'^{-k}\lambda (N) = b'^{k}\lambda(N), \;\;\;  B^{-}_{k,+} = \overline{\lambda (N)}b'^{+k} = \overline{\lambda(N)}{(b'^{\dag})}^{k};
\eeq

\item [(ii)]$\varepsilon = -,$

\beq{\label{spi04}} B^{+}_{k,-} = b'^{+k}\lambda (N) =
{(b'^{\dag})}^{k} \lambda(N), \;\;\; B^{-}_{k,-} = -
\overline{\lambda (N)}b'^{-k} = -\overline{\lambda(N)}b'^{k}, \eeq \eitem where
$\lambda(N)$ is a nonvanishing complex valued
function of the number operator $N$,
$\overline{\lambda(N)}$ denotes its complex
conjugate and $k$ is a strictly positive integer, $k=1,2,\dots.$
The expressions in (i) and (ii) can be abridged as
\beq{\label{spi05}} B^{+}_{k,\varepsilon} =
b'^{-\varepsilon k}\lambda(N), \qquad
B^{-}_{k,\varepsilon} = \varepsilon \overline{\lambda(N)}b'^{\varepsilon
k}, \eeq and denoting   the generalized dimensionless SO
interaction by
\beq{\label{spi06}} V_{k,\varepsilon}
= B^{+}_{k,\varepsilon} \sigma_{-} + B^{-}_{k,\varepsilon}
\sigma_{+}, \eeq
 the relations (\ref{spi02})  take the following compact forms
\beq{\label{spi060}} H_{R} = \imath \hbar \omega_{c} \gamma V_{k,\varepsilon}, \qquad 
\qquad H_{D} = \hbar \omega_{c} \beta V_{k,\varepsilon}. \eeq
Recall that in the study of a spin-1/2 particle of charge $-e$ subjected to a spin-orbit interaction in a perpendicular 
magnetic field ${\bf B} = -B\hat{z} = \nabla \times {\bf A}$ and an electric field ${\bf E} = E\hat{y}$ along the $y$ axis 
{\cite{shen}}, using the introduced lowering operator $a$, the both Rashba and Dresselhaus couplings are given by 
$V_{R} = \imath \sqrt{2}\eta_{R}(a\sigma_{-} - a^{\dag}\sigma_{+})$ and  $V_{D} = \sqrt{2}\eta_{D}(a^{\dag}\sigma_{-} + a\sigma_{+})$ 
which lead to the generalized dimensionless spin-orbit interaction established in {\cite{joben1}} as: 
$U_{k,\epsilon} = C^{+}_{k,\epsilon}\sigma_{+} + C^{-}_{k,\epsilon}\sigma_{-}$, 
where $C^{+}_{k,\epsilon}=a^{-\epsilon k}\lambda^{\epsilon}(N), \, C^{-}_{k,\epsilon}=\lambda^{-\epsilon}(N) a^{\epsilon k}$, 
with $\epsilon = \pm 1$.

From (\ref{spi06}) and (\ref{spi060}), the generalized Hamiltonians describing a $2D$ electron gas
mixing with full Rashba and Dresselhaus
 SO interactions, respectively,  and
effective Zeeman coupling in a perpendicular magnetic field
background is given in both relevant gauge symmetries by \beq
H_{{1, 2}_{SO}} &=& \frac{1}{2M}\left[\left(p_{x} \pm
\frac{eB}{2c}\right)^{2} + \left(p_y \mp \frac{eB}{2c}\right)^{2}
\right]\mp \frac{g \mu B}{2} \sigma_{z} + V_{SO}, 
\eeq
where $V_{SO}$ stands for the generalized Rashba or Dresselhaus spin-orbit terms given in (\ref{spi060}).  
\begin{rmq}
The relevant exactly  solvable cases known in the literature
\cite{winkler} correspond to $\beta = 0$, $\gamma = 0$ and
$\gamma=\pm \beta$ and concern with the gauge in which 
the constant magnetic field points along the positive $z-$
direction. The  eigenstates and eigenvalues of the  associated
resulting Hamiltonians are
 given by the expressions:
\begin{itemize}
 \item for $\beta = 0$ and nonvanishing Rashba SO interaction term, which couples the up-spin level
$(\Phi_{n-1},0)$ to the down-spin state $(0,\Phi_{n}), n \geq 1$
\beq{\label{spo7}}
&&\psi^{+}_{n} = \left(
\begin{array}{c}
\cos \theta_{n} \Phi_{n-1} \\
\sin \theta_{n} \Phi_{n}
\end{array}
\right), \qquad \psi^{-}_{n} = \left(
\begin{array}{c}
-\sin \theta_{n} \Phi_{n-1} \\
\cos \theta_{n} \Phi_{n}
\end{array}
\right) \cr \cr && E^{\pm}_{n} = \hbar \omega_{c} n \mp
\frac{\delta}{2}\sqrt{1+ 4\gamma^{2}n
\hbar^{2}\omega^{2}_{c}/\delta^{2} }, \eeq in which $\delta =
\hbar \omega_{c}(1+2\xi)$. The mixing angle $\theta_{n}$ is given
by $\tan 2\theta_{n} = 2\sqrt{n}\gamma \hbar \omega_{c}/\delta$
and varies from $0$ (when $\gamma = 0$) to $\pi/4$ (for weak
magnetic field or strong SO interaction). Each Landau level
$\Phi_{n} \uparrow \downarrow$ splits into two levels
$\psi^{-}_{n}$ and $\psi^{+}_{n+1}$ with the splitting gap $\Delta
= E^{-}_{n}- E^{+}_{n+1}$,
\item for $\gamma = 0$ and nonvanishing Dresselhaus SO interaction term which couples $(\Phi_n,0)$ only to 
the  $(0,\Phi_{n-1}), n \geq 1$
\beq{\label{spo8}}
&&\psi^{+}_{n} = \left(
\begin{array}{c}
\cos \theta_{n} \Phi_{n} \\
\sin \theta_{n} \Phi_{n-1}
\end{array}
\right), \qquad  \psi^{-}_{n} = \left(
\begin{array}{c}
-\sin \theta_{n} \Phi_{n} \\
\cos \theta_{n} \Phi_{n-1}
\end{array}
\right) \cr
\cr
&& E^{\pm}_{n} =  \hbar \omega_{c} n \mp \frac{\delta}{2}\sqrt{1+ 4\beta^{2}n \hbar^{2}\omega^{2}_{c}/\delta^{2} },
\eeq
with  $\delta = \hbar \omega_{c}(-1+2\xi)$, $\tan 2\theta_{n} = 2\sqrt{n}\beta \hbar \omega_{c}/\delta$.
\end{itemize}
\end{rmq}
It is also worth noticing that for the most usual widespread
situation corresponding to $\lambda(N)= -\imath \1$, $k=1$, the  cases of
 $\gamma=\pm\beta$ and
  $|\beta|< |\gamma|$ have been
treated with or without the Zeeman coupling term, using the
perturbation theory.  For more details, see \cite{zarea}. In the
latter case, the Rashba and Dresselhaus SO interactions take the
simpler forms in the gauge $
{\bf A} = \left(-\frac{B}{2}y,
\frac{B}{2}x \right)$, where $H_{R}$ and $H_{D}$ are obtained from (\ref{spi03}) and (\ref{spi04}) by
\beq{\label{so1}}
H_{R} = \imath \hbar \omega_{c} \gamma (-\imath b'^{\dag} \1 \sigma_{-} - \imath b' \1 \sigma_{+}) 
= \hbar \omega_{c} \gamma (b'^{\dag}\sigma_{-} + b'\sigma_{+}) = \hbar \omega_{c}\gamma \left(
\begin{array}{cc}
0 & b' \\
b'^{\dag} & 0 \\
\end{array}
\right),
\eeq

\beq{\label{so2}}
H_{D} = \hbar \omega_{c} \beta (\imath b'^{\dag} \1 \sigma_{+} - \imath b' \1 \sigma_{-}) 
= \hbar \omega_{c} \beta (\imath b'^{\dag}\sigma_{+} -\imath  b'\sigma_{-}) = \hbar \omega_{c}\beta \left(
\begin{array}{cc}
0 & \imath b'^{\dag} \\
-\imath b' & 0 \\
\end{array}
\right),
\eeq
respectively.
$\gamma$ and $\beta$ are dimensionless parameters defined by
$\gamma = \gamma_{0}\sqrt{\frac{2M}{\hbar^{3} \omega_{c}}}$
and $\beta = \beta_{0}\sqrt{\frac{2M}{\hbar^{3} \omega_{c}}}$.

In the sequel, we turn back to the more general situation when the interaction terms (\ref{so1}) and  (\ref{so2}) are coupled with the
generalized Hamiltonians $H_{1_{0}}$ and $H_{2_{0}}$ investigated in the previous section, describing a 2D electron in electromagnetic field
 with a Zeeman coupling term.

  Analog expressions are readily obtained for $H_{2_{SO}} = H_{2_{0}} +  V_{SO}$ in
 the second  gauge, when the constant
magnetic field points along the negative $z-$ direction,
  by replacing  $\delta$ by $\delta'$ defined with
$\xi'$.

The spinor states are formulated as
\begin{itemize}
\item for $\beta=0$ and $\gamma\neq 0$
\beq{\label{spo10}}
\psi^{+}_{n} &=& \cos \theta_{n} \left(
\begin{array}{c}
\Phi_{n-1} \\
0
\end{array}
\right) \oplus \sin \theta_{n} \left(
\begin{array}{c}
0 \\
\Phi_{n}
\end{array}
\right) \crcr
&=& \cos \theta_{n} (\chi^{+} \otimes \Phi_{n-1}) \oplus \sin \theta_{n} (\chi^{-} \otimes \Phi_{n}), \cr
\cr
\cr
\psi^{-}_{n} &=& -\sin \theta_{n} \left(
\begin{array}{c}
\Phi_{n-1} \\
0
\end{array}
\right) \oplus \cos \theta_{n} \left(
\begin{array}{c}
0 \\
\Phi_{n}
\end{array}
\right)  \crcr &=& -\sin \theta_{n} (\chi^{+} \otimes
\Phi_{n-1}) \oplus \cos \theta_{n} (\chi^{-} \otimes
\Phi_{n}), \eeq

\item for $\beta\neq0$ and $\gamma= 0$
\beq{\label{spo11}}
\psi^{+}_{n} &=& \cos \theta_{n} \left(
\begin{array}{c}
\Phi_{n} \\
0
\end{array}
\right) \oplus \sin \theta_{n} \left(
\begin{array}{c}
0 \\
\Phi_{n-1}
\end{array}
\right)\crcr
  &=& \cos \theta_{n} (\chi^{+} \otimes \Phi_{n}) \oplus \sin \theta_{n} (\chi^{-} \otimes \Phi_{n-1}),
\cr
\cr
\cr
\psi^{-}_{n} &=& -\sin \theta_{n} \left(
\begin{array}{c}
\Phi_{n} \\
0
\end{array}
\right) \oplus \cos \theta_{n} \left(
\begin{array}{c}
0 \\
\Phi_{n-1}
\end{array}
\right)\crcr
  &=& -\sin \theta_{n} (\chi^{+} \otimes \Phi_{n}) \oplus \cos \theta_{n} (\chi^{-} \otimes \Phi_{n-1}).
\eeq
\end{itemize}

In the weak Rashba SO coupling limit case ($\gamma$ negligible), we can
expand $\mathcal E^{\pm}_{n}$ using the approximation $(1+
\varepsilon)^{1/2}\simeq 1+ \varepsilon /2,  \varepsilon \ll 1$ and,  by taking $\hbar = 1$, 
 get from (\ref{spo7})
\beq{\label{relat}}
\mathcal E^{\pm}_{n} = 
\omega_{\pm}(\gamma)n, \qquad \omega_{\pm}(\gamma)= \omega_{c}\left(1 \mp
\frac{2 \gamma^{2}}{2-gc}\right)
\eeq

where  $gc \ll 2$ is assumed.  Define
\beq
\rho_{\pm}(n):=
\mathcal E^{\pm}_{1}\mathcal E^{\pm}_{2}
\cdots \mathcal E^{\pm}_{n}
\eeq
with the increasing
order $\mathcal E^{\pm}_{1} < \mathcal
E^{\pm}_{2}< \cdots < \mathcal
E^{\pm}_{n}$,  such that
\beq \rho_{\pm} (n)&=&
\prod_{q=1}^{n} \omega_{\pm}(\gamma)q  = n! 
(\omega_{\pm}(\gamma))^{n}.
\eeq
.

\subsection{Spectral decomposition}

Introduce the passage operators from the Hilbert space $\mathcal S$ with eigenbasis 
$\{|\psi^{\pm}_{n}\rangle \}$ to $\{|\chi^{\pm} \otimes \Phi_{n}\rangle \}$ and vice versa, 
$\chi^{\pm}$ being  the natural basis of $\C^{2}$,
 given by

\beq
\mathcal U|\chi^{\pm} \otimes \Phi_{n}\rangle = |\psi^{\pm}_{n}\rangle \qquad 
\mathcal U^{\dag}|\psi^{\pm}_{n}\rangle = |\chi^{\pm} \otimes \Phi_{n}\rangle    
\eeq

where $\mathcal U, \mathcal U^{\dag}$ expand as  

\beq
\mathcal U = \sum_{n = 0,\pm}^{\infty}  |\psi^{\pm}_{n}\rangle \langle \chi^{\pm} \otimes \Phi_{n}| 
\qquad \mathcal U^{\dag} = \sum_{n = 0,\pm}^{\infty} |\chi^{\pm} \otimes \Phi_{n}\rangle  \langle \psi^{\pm}_{n}| 
\eeq

respectively. $\mathcal U, \mathcal U^{\dag}$ are obtained as mutually adjoint through  the following identities:

\beq
\mathcal U^{\dag}\mathcal U  = \sum_{n = 0,\pm}^{\infty} |\chi^{\pm} \otimes \Phi_{n}\rangle \langle \chi^{\pm} \otimes \Phi_{n} |
=  \mathbb I_{2} \otimes I_{\mathfrak H},
\qquad  \mathcal U \mathcal U^{\dag} = \sum_{n = 0,\pm}^{\infty} |\psi^{\pm}_{n}\rangle  \langle \psi^{\pm}_{n} | =  \mathbb I_{\mathcal S},
\eeq

where $\mathbb I_{2} \otimes I_{\mathfrak H}$ and $\mathbb I_{\mathcal S}$ are the  identity  on $\C^{2} \otimes \mathfrak H$ and 
$\mathcal S$, respectively.
Then, the Hamiltonians $H_{{1, 2}_{SO}}$ can be rewritten in a diagonal form as below:

\beq{\label{hq00}}
\mathbb H^{red}_{{1, 2}_{SO}}  =   \mathcal U^{\dag} H_{{1, 2}_{SO}} \mathcal U  =  
\sum_{n = 0,\pm}^{\infty} |\chi^{\pm} \otimes \Phi_{n}\rangle \mathcal E^{\pm}_{n} \langle \chi^{\pm} \otimes \Phi_{n} |.
\eeq

\section{Canonical QVCS and VCS }

In this section,  we first  provide   the canonical  QVCS related to the Hamitonian   $\mathbb H^{red}_{{1}_{SO}}$ 
 and then discuss the VCS  construction with their  connection with the QVCS. Then,   
statistical properties of the  constructed QVCS are  investigated  through studying some expectations and the uncertainty relation.

\subsection{Canonical QVCS }

The QVCS  are built using the complex representation of quaternions by $2 \times 2$ matrices \cite{thirulogasanthar-ali}. 
Using the basis matrices,
\beq
\sigma_{0} =  \left(
\begin{array}{cc}
1 & 0 \\
0 & 1 
\end{array}
\right), \; \imath \sigma_{1} =  \left(
\begin{array}{cc}
0 & \imath \\
\imath  & 0 
\end{array}
\right), \; -\imath \sigma_{2} =  \left(
\begin{array}{cc}
0 & -1 \\
1 & 0 
\end{array}
\right), \; \imath \sigma_{3} =  \left(
\begin{array}{cc}
\imath  & 0 \\
0 & -\imath 
\end{array}
\right),
\eeq
where $\sigma_{1}, \sigma_{2}$ and $\sigma_{3}$ are the usual Pauli matrices, a general quaternion is written as
\beq{\label{quater00}}
q = x_{0}\sigma_{0} + \imath \underline{x}. \underline{\sigma}
\eeq
with $x_{0} \in \R, \,  \underline{x} = (x_{1},x_{2},x_{3}) \in \R^{3}$ and $\underline{\sigma} = (\sigma_{1}, -\sigma_{2}, \sigma_{3})$. Thus,

\beq
\mathfrak q =  \left(
\begin{array}{cc}
x_{0} + \imath x_{3} & -x_{2} + \imath x_{1} \\
x_{2} + \imath x_{1}  & x_{0} - \imath x_{3} 
\end{array}
\right).
\eeq

Introducing the polar coordinates as:

$x_{0} = r\cos{\eta}, \; x_{1} = r\sin{\eta} \sin{\phi} \cos{\psi}, \; x_{2} = r \sin{\eta} \sin{\phi} \sin{\psi}, \; 
x_{3} = r\sin{\eta}\cos{\phi}$,
where  $r \in [0, \infty), \phi \in [0,\pi]$ and $\eta, \psi \in [0, 2\pi]$, it follows that
\beq{\label{qvcs_00}}
\mathfrak q = A(r)e^{\imath \eta \sigma(\hat n)}
\eeq
where
\beq{\label{quater01}}
A(r) = r \sigma_{0}, \qquad \sigma (\hat n)  = \left(
\begin{array}{cc}
\cos{\phi} & e^{\imath \psi}\sin{\phi} \\
e^{-\imath \psi}\sin{\phi} & -\cos{\phi}
\end{array}
\right) \qquad \mbox{and} \qquad \sigma (\hat n)^{2} = \sigma_{0} = \mathbb I_{2}.
\eeq

The field of quaternions is denoted by $\mathbb H$, this algebra being generated by $\{1,\hat i, \hat j, \hat k\}$, where $\hat i, \hat j, \hat k$ 
are imaginary units. 
The real quaternions denoted by 
$Q_{\R}$ are given by
\beq
Q_{\R} = \{q = x_{0} + x_{1}\hat i + x_{2}\hat j + x_{3}\hat k| \; \hat i^{2} = \hat j^{2} = \hat k^{2} = -1
,\,  \hat i \hat j \hat k = -1, \,  x_{0}, x_{1}, x_{2},x_{3} \in \R \},
\eeq 
where $q$ can be written as $q  = q(x_{0}, \underline{x}) = x_{0} + \underline{i}. \underline{x}$, with $\underline{i} = (\hat i,  \hat j,  \hat k)$ 
and $\underline{x} = (x_{1}, x_{2},x_{3})$ given as in (\ref{quater00}). The complex quaternions denoted by $Q_{\C}$ are given by
$q = x_{0} + \imath x_{1}\hat i + \imath x_{2}\hat j + \imath x_{3}\hat k =  x_{0} + \imath \underline{i}. \underline{x}$.

Remark that in the definition of the Hamiltonian given in (\ref{spo3}), the harmonic oscillator part of ${H}_{1_0}$, given by 
$\hbar \omega_{c} \left(b'^{\dag}b' + \frac{\1}{2}\right)\sigma_{0}$, can be rewritten 
in the supersymmetric  quantum mechanics as 
\beq
H^{SUSY} = \left(
\begin{array}{cc}
H^{b} & 0 \\
0 &  H^{f}
\end{array}
\right) = \hbar \omega_{c}\left(
\begin{array}{cc}
b'^{\dag}b'  & 0 \\
0 &  b'b'^{\dag}
\end{array}
\right),
\eeq
where the supersymmetric partner Hamiltonians, $H^{b}, H^{f}$ are given by 
 $H^{b} = H_{1_{OSC}} - \frac{\hbar \omega_{c}}{2}\1$ and $H^{f} = H_{1_{OSC}} + \frac{\hbar \omega_{c}}{2}\1$, 
with $H_{1_{OSC}} = \hbar \omega_{c} \left(b'^{\dag}b' + \frac{\1}{2}\right)$, respectively. 
This Hamiltonian can also be rewritten as $H^{SUSY} = \{Q, Q^{\dag}\}$, with $Q = \sqrt{\hbar \omega_{c}}\left(
\begin{array}{cc}
 0 & 0 \\
b' & 0
\end{array}
\right)=  \sqrt{\hbar \omega_{c}}\, b'\sigma_{-}$
and $Q^{\dag}$ the supercharges. 
The annihilation operator defined on $\C^{2}\otimes \mathfrak H$ by
\beq{\label{anihil}}
\mathcal A' = \left(
\begin{array}{cc}
 b' & 0 \\
0 & b'
\end{array}
\right)
\eeq
satisfies the commutation relation $[H^{SUSY}, \mathcal A'] = -\hbar \omega_{c}\mathcal A'$. The connection with the supersymmetric harmonic 
oscillator in this study appears  by obtaining the QVCS  denoted by $|\mathfrak q;\pm\rangle$, known in the literature as the 
{\it{quaternionic canonical coherent states}} \cite{thirulogasanthar-ali}, given on $\C^{2}\otimes \mathfrak H$ as follows
\beq{\label{cqvcs}}
|\mathfrak q;\pm\rangle
&=& \frac{e^{-\frac{r^{2}}{2}}}{\sqrt{2}}\sum_{n=0}^{\infty}\frac{\mathfrak q^{n}}{\sqrt{n !}} |\chi^{\pm} \otimes \Phi_{n}\rangle,
\eeq
where $\mathfrak q$ is given in (\ref{qvcs_00}), as the eigenstates of the operator $\mathcal A'$ in (\ref{anihil}).

The  QVCS  related to the Hamiltonian $\mathbb H^{red}_{{1}_{SO}}$ can be defined
as
\beq{\label{spo12}}
|\mathfrak q;\pm\rangle_{1}
&=& \left(\mathcal N_{(\star)}(|\mathfrak q|)\right)^{-1/2}\sum_{n=0}^{\infty}\frac{\mathfrak q^{n}}{\sqrt{\rho_{(\star)}(n)}}
|\chi^{\pm} \otimes \Phi_{n}\rangle, 
\eeq
where $\mathcal N_{(\star)}(|\mathfrak q|),  \rho_{(\star)}(n)$ stand for $\mathcal N_{+}(|\mathfrak q|)$ and  $\rho_{+}(n)$ 
(resp. $\mathcal N_{-}(|\mathfrak q|)$ and  $\rho_{-}(n)$). 
The variables $\mathfrak q$ are as in (\ref{qvcs_00}). The $2 \times 2$ matrix-valued functions $A(r), \sigma(\hat{n})$  satisfy the following 
properties \cite{thirulogasanthar-ali} 
(assumed to hold almost everywhere with respect to appropriate measures):
\beq{\label{prop1}}
\sigma(\hat{n}) = \sigma(\hat{n})^{\dag}\qquad  [A(r),A(r)^{\dag}] = 0 \qquad  [A(r), \sigma(\hat{n})] = 0.
\eeq

The normalization constants $\mathcal N_{(\star)}(|\mathfrak q|)$  ensured by the relations
\beq
\sum_{\pm}\,_{1}\langle \mathfrak q;\pm|\mathfrak q;\pm\rangle_{1} = 1,
\eeq
are given, with respect to (\ref{prop1}) for $|\mathfrak q;\pm\rangle_{1}$
by
\beq{\label{spo14}}
\mathcal N_{(\star)}(|\mathfrak q|)  =  \mathcal N_{(\star)}(r)
&=& 2\sum_{n=0}^{\infty}\frac{r^{2n}}{n !(\omega_{(\star)}(\gamma))^{n}} = 2e^{\frac{r^{2}}{\omega_{(\star)}(\gamma)}}.
\eeq

The  $\mathbb H^{red}_{{1}_{SO}}$ QVCS  (\ref{spo12})
satisfy on the Hilbert space  $\C^{2} \otimes \mathfrak H$  the following resolution of the identity
\beq{\label{qvcsresolu}}
\sum_{\pm}\int_{\mathcal D}|\mathfrak q;\pm\rangle_{1} \,
W_{(\star)}(r)\, _{1}\langle \mathfrak q;\pm|\,d\nu &=& \mathbb I_{2} \otimes I_{\mathfrak H},
\eeq
 
$W_{(\star)}(r)$ is the density function and $\mathcal D =
\R^{+}
\times
[0,2\pi) \times S^{2} $,
where $S^{2}$ is the surface of the unit two-sphere on which the points  with coordinates
 $(\phi,\psi)$ are  defined.
The measure $d\nu$ is such that
$d\nu(r,\eta,\phi,\psi) = rdrd\eta d\Omega (\phi,\psi)$,
with $d\Omega (\phi,\psi) = \frac{1}{4\pi}\sin \phi d\phi d\psi$.

The relation (\ref{qvcsresolu})
translates in the Stieljes moment problems
\beq
\int_{0}^{\infty}r^{2n + 1}\varrho_{(\star)}(r) dr = \rho_{(\star)}(n),
\eeq

which are solved by 

\beq{\label{dens00}}
\varrho_{(\star)}(r) = \frac{2}{\omega_{(\star)}(\gamma)}e^{-\frac{r^{2}}{\omega_{(\star)}(\gamma)}}
\eeq

where the density functions are given by
\beq{\label{dens01}}
 W_{(\star)}(r) = \frac{\mathcal N_{(\star)}(r)\varrho_{(\star)}(r)}{2\pi}.
\eeq

Treating the vectors $|\mathfrak q;+\rangle_{1}$ and $|\mathfrak q;-\rangle_{1}$ as elements of a basis, we
shall define a general QVCS for $\mathbb H^{red}_{{1}_{SO}}$ (resp. $\mathbb H^{red}_{{2}_{SO}}$) as a linear combination,
\beq
|\mathfrak q, \chi \rangle =  c_{+}|\mathfrak q;+\rangle_{1} + c_{-}|\mathfrak q;-\rangle_{1}
\eeq
where $c_{\pm} \in \C, |c_{+}|^{2} + |c_{-}|^{2} = 1$ and $\chi = \sum_{\pm}c_{\pm}\chi^{\pm}$.

It is worth noticing that
the constructed QVCS  are related to the Weyl-Heisenberg group $G_{W-H}$
as observed in \cite{thirulogasanthar-ali}.
Indeed, from the relations
\beq
e^{\mathfrak q \otimes b'^{\dag} - \mathfrak q^{\dag} \otimes b'} = e^{(-1/2)[\mathfrak q \otimes b'^{\dag},\, -\mathfrak q^{\dag} \otimes b']}\,
e^{\mathfrak q \otimes b'^{\dag}} e^{-\mathfrak q^{\dag}\otimes b'}, \quad
  [\mathfrak q^{\dag} \otimes b', \mathfrak q \otimes b'^{\dag}] = r^{2}\mathbb I_{2}\otimes I_{\mathfrak H}
\eeq

since
\beq
e^{-\mathfrak q \otimes b'}|\chi^{\pm}\otimes \Phi_{0}\rangle = |\chi^{\pm} \otimes \Phi_{0}\rangle, 
\eeq
we can write
\beq{\label{spo22}}
\frac{1}{\sqrt{2}}e^{\frac{1}{\omega_{(\star)}(\cdot)}[\mathfrak q \otimes b'^{\dag} - \mathfrak q^{\dag} \otimes b']}
|\chi^{\pm}\otimes \Phi_{0}\rangle 
&=& \frac{e^{-\frac{r^{2}}{2\omega_{(\star)}(\cdot)}}}{\sqrt{2}}\sum_{n=0}^{\infty}
\frac{\mathfrak q^{n}}{\sqrt{n !(\omega_{(\star)}(\cdot))^n}}|\chi^{\pm} \otimes \Phi_{n}\rangle,  
\eeq
where $\omega_{(\star)}(\cdot)$ stands for $\omega_{(\star)}(\beta)$ and $\omega_{(\star)}(\gamma)$, respectively.
Therefore, in the case of the Hamiltonian $H_{1_{SO}}$,
the QVCS   can be rewritten as
\beq
|\mathfrak q;\pm\rangle_{1} =  \frac{1}{\sqrt{2}}\,
U^{(\star)}_{1}(0, \mathfrak q)|\chi^{\pm}\otimes \Phi_{0}\rangle
\qquad |\mathfrak q;\pm\rangle^{\sim}_{1} = \frac{1}{\sqrt{2}}
\,
\tilde{U}^{(\star)}_{1}(0, \mathfrak q)|\chi^{\pm} \otimes \Phi_{0}\rangle
,
\eeq
in analogy with the case of the canonical coherent states \cite{ali-antoine-gazeau}, where the operators $b'$
and ${b'^{\dag}}$ are identical to those given in (\ref{es11}).

The unitary operator $\tilde{U}$ has the form \cite{thirulogasanthar-ali}
\beq
\tilde{U}(0,\mathfrak q) = e^{\mathfrak q \otimes b'^{\dag} - \mathfrak q^{\dag} \otimes b'} =
u(\eta,\phi)\left(
\begin{array}{cc}
U(0,q,p) & 0 \\
0 & U(0,q,-p)
\end{array}
\right) u(\eta,\phi)^{\dag},
\eeq
and for fixed $(\eta,\phi), \eta, \phi, \in [0,\pi]$ and $p,q,$ such that $z = \frac{q-\imath p}{\sqrt{2}}$,
 the operators $\tilde{U}^{(\star)}_{1}(0, \mathfrak q)$ and $\tilde{U}^{(\star)}_{2}(0,
\mathfrak q)$ are defined for Rashba and Dresselhaus SO interactions, respectively,  by
 \beq
&& U^{(\star)}_{1}(0,\mathfrak q) = e^{\frac{1}{\omega_{(\star)}(\gamma)}[\mathfrak q \otimes b'^{\dag} -
 \mathfrak q^{\dag} \otimes b']}, \qquad U^{(\star)}_{2}(0,\mathfrak q) =
e^{\frac{1}{\omega_{(\star)}(\gamma)}[\mathfrak
q \otimes \hat d'^{\dag} -
\mathfrak q^{\dag} \otimes \hat d']},
\eeq
\beq
&& \tilde{U}^{(\star)}_{1}(0,\mathfrak q) = e^{\frac{1}{\omega_{(\star)}(\beta)}[\mathfrak q \otimes b'^{\dag} -
 \mathfrak q^{\dag} \otimes b']}, \qquad \tilde{U}^{(\star)}_{2}(0,\mathfrak q) =
e^{\frac{1}{\omega_{(\star)}(\beta)}[\mathfrak
q \otimes \hat d'^{\dag} -
\mathfrak q^{\dag} \otimes \hat d']}.
\eeq
They  realize unitary (reducible) representations of the Weyl-Heisenberg group $G_{W-H}$
\cite{thirulogasanthar-ali} on the Hilbert space $\C^{2} \otimes \mathfrak H$.

\subsection{Vector coherent states related to  $\mathbb H^{red}_{{1}_{SO}}$}

This paragraph deals with the Hamiltonian  $\mathbb H^{red}_{{1}_{SO}}$  VCS and QVCS construction. Each set of vectors satisfies a 
resolution of the identity on the Hilbert space $\C^{2} \otimes \mathfrak H$.

The VCS are given on $\C^{2} \otimes \mathfrak H$ as 

\beq
|Z;\pm\rangle_{1} = (\mathcal N(Z))^{-1/2}  
\sum_{n=0}^{\infty} (R(n))^{-1/2}
 Z^{n}  |\chi^{\pm} \otimes \Phi_{n}\rangle
\eeq
where $Z = \mbox{diag}(z_{1}, z_{2}), z_{1} = r_{1}e^{i \varphi_{1}},  z_{2} = r_{2}e^{i \varphi_{2}}, r_{1}, r_{2}  \in [0, \infty),  
\varphi_{1}, \varphi_{2} \in [0, 2\pi)$ and 
$R(n) = \mbox{diag}(\rho_{+}(n), \rho_{-}(n))$.

In this case,  the normalization factor is given by

\beq
\mathcal N(Z) = e^{r^{2}_{1}/\omega_{+}(\gamma)} + e^{r^{2}_{2}/\omega_{-}(\gamma)}.
\eeq

They fulfill  on $\C^{2} \otimes \mathfrak H$ the following resolution of the identity 

\beq
\sum_{\pm}\int_{\mathcal D_{1} \times \mathcal D_{2}}|Z;\pm\rangle_{1} W(r_{1}, r_{2})\, _{1}\langle Z;\pm|d\mu = \mathbb I_{2}\otimes I_{\mathfrak H}
\eeq

where the measure  $d\mu$ is given on $\mathcal D_{1} \times \mathcal D_{2} = \{[0, \infty) \times [0, 2\pi)\}^{2}$  by 
$d\mu(r_{1}, r_{2}, \varphi_{1}, \varphi_{2}) = \frac{1}{4\pi^{2}}r_{1}dr_{1}r_{2}dr_{2}d\varphi_{1}d\varphi_{2}$. 
From (\ref{dens00}) and (\ref{dens01}), the density function is now provided by
\beq 
 W(r_{1}, r_{2}) = 
\mathcal N (Z) \frac{ 4 e^{-r^{2}_{1}/\omega_{+}(\gamma)}  
e^{-r^{2}_{2}/\omega_{-}(\gamma)} }{  \omega_{+}(\gamma)\omega_{-}(\gamma)}.
\eeq

Let $\mathcal Z = diag(z, \bar z)$ where $ z = \tilde r e^{\imath \vartheta}, \tilde r \in [0, \infty), \vartheta \in [0, 2\pi)$. 
Consider 
$\mathfrak Z = U \mathcal Z U^{\dag}, U \in SU(2)$ with  
$U = u_{\phi_{1}} u_{\varphi} u_{\phi_{2}}, \, \phi_{1}, \phi_{2} \in [0, 2\pi) $ 
and 
$\varphi \in [0, \pi]$,  where 

\beq
u_{\phi_{j}} = \left(\begin{array}{cc}
e^{\imath \frac{\phi_{j}}{2}} & 0 \\
0 & e^{-\imath \frac{\phi_{j}}{2}}
                     \end{array}
\right),  \, j=1,2, 
\quad  
u_{\varphi} = \left(\begin{array}{cc}
\cos{\frac{\varphi}{2}} & \imath \sin{\frac{\varphi}{2}} \\
\imath \sin{\frac{\varphi}{2}} & \cos{\frac{\varphi}{2}}
                     \end{array}
\right).
\eeq

Introduce the quaternion $\mathfrak Q = B(\tilde r)e^{\imath \vartheta \tilde \sigma(\hat k) }$ with $B(\tilde r) = \tilde r \mathbb I_{2}$  and 

\beq
\tilde \sigma(\hat k) = \left(\begin{array}{cc}
\cos{\varphi} & e^{\imath \varrho}\sin{\varphi} \\
e^{-\imath \varrho}\sin{\varphi}  & -\cos{\varphi}
                                             \end{array}
\right)
\eeq

  and   $\varphi \in [0,\pi], 
 \vartheta, \varrho \in [0,2\pi].$ For $\phi_{1} = \phi_{2} = \varrho$  we get 
$\mathfrak Z = \tilde r(\mathbb I_{2}\cos{\vartheta} + \imath \tilde \sigma(\hat k)\sin{\vartheta})= \mathfrak Q$. 
Then,  the QVCS are given by 
\beq{\label{spoqvcs23}}
|\mathfrak Q;\pm\rangle_{1} = (\mathcal N(\tilde r))^{-1/2}\sum_{n=0}^{\infty} (R(n))^{-1/2}
 \mathfrak Q^{n}  |\chi^{\pm} \otimes \Phi_{n}\rangle.
\eeq 

The normalization factor is provided by 

\beq
\mathcal N(\tilde r) = e^{\tilde r^{2}/\omega_{+}(\gamma)} + e^{\tilde r^{2}/\omega_{-}(\gamma)}.
\eeq

They satisfy on $\C^{2} \otimes \mathfrak H$ the following resolution of the identity 

\beq
\sum_{\pm}\int_{\mathcal D }|\mathfrak Q;\pm\rangle_{1} W(\tilde r) \, _{1}\langle \mathfrak Q;\pm|d\mu 
= \mathbb I_{2}\otimes I_{\mathfrak H}
\eeq

where   $\mathcal D = [0, \infty)   \times [0, 2\pi) \times S^{2}$ 
and $d\mu = \frac{1}{4\pi} \tilde r d\tilde r \sin{\varphi} d\varphi d\varrho d\vartheta$ with 

\beq
W(\tilde r) =  \left(\begin{array}{cc}
 \frac{\mathcal N(\tilde r)}{\pi}\frac{(2\omega_{c})^{n+1}}{(\omega_{-}(\gamma))^n}
\frac{  e^{-\tilde r^{2}\left[\frac{2\omega_{c}}{\omega_{+}(\gamma)\omega_{-}(\gamma)} 
\right]}}{\omega_{+}(\gamma)\omega_{-}(\gamma)} & 0 \\
0 &  \frac{\mathcal N(\tilde r)}{\pi}\frac{(2\omega_{c})^{n+1}}{(\omega_{+}(\gamma))^n}
\frac{  e^{-\tilde r^{2}\left[\frac{2\omega_{c}}{\omega_{+}(\gamma)\omega_{-}(\gamma)} 
\right]}}{\omega_{+}(\gamma)\omega_{-}(\gamma)}
                \end{array}
\right).
\eeq

{\bf Proof.}

\beq
&&\sum_{\pm}\int_{\mathcal D }|\mathfrak Q;\pm\rangle_{1}  W(\tilde r) \, _{1}\langle \mathfrak Q;\pm|d\mu \cr
&&= \int_{ \mathcal D }W(\tilde r) \times \cr
 &&\sum_{\pm}
\left| (\mathcal N(\tilde r))^{-1/2} \sum^{\infty}_{n=0}\frac{\mathfrak Q^{n}\chi^{\pm} \otimes \Phi_{n}}{\sqrt{R(n)}} \right\rangle
\left\langle (\mathcal N(\tilde r))^{-1/2} \sum^{\infty}_{m=0}\frac{\mathfrak Q^{m}\chi^{\pm} \otimes \Phi_{m}}{\sqrt{R(m)}} \right|d\mu \cr
&&= \sum_{\pm}\sum^{\infty}_{m=0}\sum^{\infty}_{n=0} \int_{ \mathcal D }
\frac{ W(\tilde r)}{\mathcal N(\tilde r)\sqrt{R(m)R(n)}}\mathfrak Q^{n}
|\chi^{\pm}\rangle \langle \chi^{\pm}|  {\mathfrak Q^{\dag}}^{m} \otimes |\Phi_{n}\rangle \langle \Phi_{m}| d\mu \cr
&&= \sum^{\infty}_{m=0}\sum^{\infty}_{n=0} \int_{ \mathcal D}
\frac{W(\tilde r)}{\mathcal N(\tilde r)\sqrt{R(m)R(n)}}B(\tilde r)^{n}e^{i n\vartheta \tilde \sigma(\hat k)}\left(\sum_{\pm}
|\chi^{\pm}\rangle \langle \chi^{\pm}|\right) \cr
&& \times B(\tilde r)^{m\dag}e^{-i m \vartheta \tilde \sigma(\hat k)\dag} \otimes |\Phi_{n}\rangle \langle \Phi_{m}| d\mu. \nonumber
\eeq
Using the facts
\beq{\label{propmat}}
&&\sum_{\pm}
|\chi^{\pm}\rangle \langle \chi^{\pm}| = \mathbb I_{2}, \cr
&&\tilde \sigma(\hat k)^{\dag} = \tilde \sigma(\hat k) \qquad  \mbox{and} \cr 
&& 
\int_{0}^{2\pi}\int_{0}^{2\pi}\int_{0}^{\pi}e^{i(n-m)\vartheta\tilde \sigma(\hat k)}
\sin{\varphi}d\varphi d\varrho d\vartheta = 
\left\{
              \begin{array}{lll}
              0 \quad \mbox{if}  \quad m \neq n, \\
               \\
              8\pi^{2} \mathbb I_{2}
\quad \mbox{if}  \quad m = n,
               \end{array}
\right.
\eeq
 
the last line is written as
\beq{\label{line}}
&& \sum^{\infty}_{m=0}\sum^{\infty}_{n=0} \int_{ \mathcal D}
\frac{W(\tilde r)}{\mathcal N(\tilde r)\sqrt{R(m)R(n)}}B(\tilde r)^{n}e^{i n\vartheta \tilde \sigma(\hat k)}\left(\sum_{\pm}
|\chi^{\pm}\rangle \langle \chi^{\pm}|\right) \cr
&& \times B(\tilde r)^{m\dag}e^{-i m \vartheta \tilde \sigma(\hat k)\dag} \otimes |\Phi_{n}\rangle \langle \Phi_{m}| d\mu\cr
&& =\sum^{\infty}_{n=0}\int_{0}^{\infty}\frac{2\pi W(\tilde r)}{\mathcal N(\tilde r)}  |B(\tilde r)|^{2n}{\mbox{diag}}
\left(\frac{1}{\rho_{+}(n)}, \frac{1}{\rho_{-}(n)} \right)
\otimes |\Phi_{n}\rangle \langle \Phi_{m}| \tilde r d\tilde r.
\eeq

Taking
\beq
W(\tilde r) =  \left(\begin{array}{cc}
 \frac{\mathcal N(\tilde r)}{\pi}\frac{(2\omega_{c})^{n+1}}{(\omega_{-}(\gamma))^n}
\frac{  e^{-\tilde r^{2}\left[\frac{2\omega_{c}}{\omega_{+}(\gamma)\omega_{-}(\gamma)} 
\right]}}{\omega_{+}(\gamma)\omega_{-}(\gamma)} & 0 \\
0 &  \frac{\mathcal N(\tilde r)}{\pi}\frac{(2\omega_{c})^{n+1}}{(\omega_{+}(\gamma))^n}
\frac{  e^{-\tilde r^{2}\left[\frac{2\omega_{c}}{\omega_{+}(\gamma)\omega_{-}(\gamma)} 
\right]}}{\omega_{+}(\gamma)\omega_{-}(\gamma)}
                \end{array}
\right)
\eeq

we get from (\ref{line}) that
\beq
&&\sum_{\pm}\int_{\mathcal D }|\mathfrak Q;\pm\rangle_{1}  W(\tilde r)\, _{1}\langle \mathfrak Q;\pm|d\mu \cr
& = &   \sum^{\infty}_{n=0}\left(\begin{array}{cc}
\frac{(2\omega_{c})^{n+1}}{(\omega_{-}(\gamma))^n}
\int_{0}^{\infty}\frac{2 \tilde r^{2n}}{\rho_{+}(n)} \frac{e^{-\tilde r^{2}\left[\frac{2\omega_{c}}{\omega_{+}(\gamma)\omega_{-}(\gamma)} 
\right]}}{\omega_{+}(\gamma)\omega_{-}(\gamma)}\tilde rd\tilde r
  & 0 \\
0 & \frac{(2\omega_{c})^{n+1}}{(\omega_{+}(\gamma))^n}
\int_{0}^{\infty}\frac{2 \tilde r^{2n}}{\rho_{-}(n)}  \frac{e^{-\tilde r^{2}\left[\frac{2\omega_{c}}{\omega_{+}(\gamma)\omega_{-}(\gamma)} 
\right]}}{\omega_{+}(\gamma)\omega_{-}(\gamma)}\tilde rd\tilde r
    \end{array}
\right)  \cr
&& \times \mathbb I_{2} \otimes |\Phi_{n}\rangle \langle \Phi_{n}| \cr
&=&
\mathbb I_{2} \otimes \sum_{n=0}^{\infty}|\Phi_{n}\rangle \langle \Phi_{n}| = \mathbb I_{2}\otimes I_{\mathfrak H} \nonumber
\eeq

where the following moment problems 

\beq
\frac{(2\omega_{c})^{n+1}}{(\omega_{\mp}(\gamma))^n}
\int_{0}^{\infty}\frac{2 \tilde r^{2n}}{\rho_{\pm}(n)} \frac{e^{-\tilde r^{2}\left[\frac{2\omega_{c}}{\omega_{+}(\gamma)\omega_{-}(\gamma)} 
\right]}}{\omega_{+}(\gamma)\omega_{-}(\gamma)}\tilde rd\tilde r = 1
\eeq

are satisfied.

$\hfill{\square}$

\subsection{Physical properties}
  In this section, we study the physical properties of the  QVCS  (\ref{spo12}) and (\ref{spoqvcs23}). 
The generalized annihilation, creation and number operators                                                                           
   are introduced and allow the definitions of the quadrature 
operators. Their actions on the QVCS  (\ref{spo12}) and (\ref{spoqvcs23}) are then derived. Next, the expectations of these operators are given in 
the second paragraph. The last paragraph discusses the time evolution of the  constructed QVCS.
\subsubsection{Generalized annihilation, creation and number operators}
The generalized
annihilation, creation and number operators, given on the Hilbert space $\mathfrak H$, with respect to the basis $\{|\Phi_{n}\rangle\}^{\infty}_{n=0}$ 
 can  be defined as:
\beq
a|\Phi_{n}\rangle  = \sqrt{x_{n}}|\Phi_{n-1}\rangle, \quad  a^{\dag}|\Phi_{n}\rangle = \sqrt{x_{n+1}}|\Phi_{n+1}\rangle, \quad  
N'|\Phi_{n}\rangle = x_{n}|\Phi_{n}\rangle,
\eeq
with $x_{n} = n\omega_{(\star)}(\cdot)$ such that $x_{n} != n !(\omega_{(\star)}(\cdot))^{n}, x_{0} ! = 1$.

The corresponding operators for the QVCS  (\ref{spo12}) are given on $\C^{2} \otimes \mathfrak H$ by

\beq{\label{op00}}
A = \mathbb I_{2} \otimes a, \quad A^{\dag} = \mathbb I_{2} \otimes a^{\dag}, \quad N = \mathbb I_{2} \otimes N'.
\eeq

Their actions on the the QVCS  (\ref{spo12}) are

\beq{\label{quad01}}
A|\mathfrak q;\pm\rangle_{1} = \mathfrak q|\mathfrak q;\pm\rangle_{1},
\eeq
\beq{\label{quad02}}
A^{\dag}|\mathfrak q;\pm\rangle_{1} = \,\left(\mathcal N_{(\star)}(|\mathfrak q|)\right)^{-1/2}
\sum_{n=0}^{\infty}\mathfrak q^{n} \sqrt{\frac{x_{n+1}}{x_{n} !}}  |\chi^{\pm} \otimes \Phi_{n+1}\rangle,
\eeq

\beq
N|\mathfrak q;\pm\rangle_{1} = \,\left(\mathcal N_{(\star)}(|\mathfrak q|)\right)^{-1/2} 
\sum_{n=0}^{\infty}\frac{ \mathfrak q^{n} x_{n}}{\sqrt{x_{n} !}} |\chi^{\pm} \otimes \Phi_{n}\rangle.
\eeq
Furthermore, we have
\beq
[A, A^{\dag}]|\mathfrak q;\pm\rangle_{1} &=& \,\left(\mathcal N_{(\star)}(|\mathfrak q|)\right)^{-1/2}
\sum_{n=0}^{\infty}\mathfrak q^{n}\frac{(x_{n+1} - x_{n})}{\sqrt{x_{n} !}} |\chi^{\pm} \otimes \Phi_{n} \rangle, \crcr
[Q,P]|\mathfrak q;\pm\rangle_{1} &=& \imath \,\left(\mathcal N_{(\star)}(|\mathfrak q|)\right)^{-1/2}
\sum_{n=0}^{\infty}\mathfrak q^{n}\frac{(x_{n+1} - x_{n})}{\sqrt{x_{n} !}} |\chi^{\pm} \otimes \Phi_{n}\rangle \crcr
&=& \imath [A, A^{\dag}]|\mathfrak q;\pm\rangle_{1}
\eeq

where 

\beq
Q = \frac{A + A^{\dag}}{\sqrt{2}}, \qquad  P = \frac{A - A^{\dag}}{\imath \sqrt{2}}.
\eeq

\subsubsection{Expectations and uncertainty relation}

The expectations of the operators $A, A^{\dag}, N, P$ and $Q$ for the QVCS  (\ref{spo12}) are derived in the quantum state 
$|\mathfrak q;\pm\rangle_{1}$
as follows:
\beq
\langle A \rangle_{|\mathfrak q;\pm\rangle_{1}} = \,  _{1}\langle \mathfrak q;\pm|A|\mathfrak q;\pm\rangle_{1} 
= \langle \chi^{\pm}|\mathfrak q\chi^{\pm} \rangle = \frac{r}{2}(\cos{(\eta)} \pm \imath \sin{(\eta)}\cos{( \phi)}),
\eeq

\beq
\langle A^{\dag} \rangle_{|\mathfrak q;\pm\rangle_{1}} 
= \,  _{1}\langle \mathfrak q;\pm|A^{\dag}|\mathfrak q;\pm\rangle_{1} 
= \langle \chi^{\pm}\mathfrak q|\mathfrak \chi^{\pm} \rangle = \frac{r}{2}(\cos{(\eta)} \mp \imath \sin{(\eta)}\cos{( \phi)}),
\eeq

\beq
\langle Q \rangle_{|\mathfrak q;\pm\rangle_{1}} 
= \frac{r}{\sqrt{2}} \cos{(\eta)}, \;\; \langle P \rangle_{|\mathfrak q;\pm\rangle_{1}}
 = \pm \frac{r}{\sqrt{2}}  \sin{(\eta)} \cos{( \phi)}, \;\; \langle N \rangle_{|\mathfrak q;\pm\rangle_{1}} = \frac{1}{2}|\mathfrak q|^{2} 
= \frac{r^{2}}{2}.
\eeq

As other property of the QVCS  (\ref{spo12}), the following uncertainty relation,
\beq
(\Delta Q)^{2} (\Delta P)^{2} \geq \frac{1}{4}\left(\left(\mathcal N_{(\star)}(|\mathfrak q|)\right)^{-1}
\sum_{n=0}^{\infty}\frac{(x_{n+1} - x_{n})}{x_{n} !}\langle \mathfrak q^{n} \chi^{\pm}|\mathfrak q^{n} \chi^{\pm}\rangle\right)^{2}
\eeq
is satisfied.
In the particular situation where $x_{n} = n,\mathcal N_{(\star)}(|\mathfrak q|) = 2e^{r^{2}}$, we readily obtain
\beq
(\Delta Q)^{2} (\Delta P)^{2} \geq  \frac{1}{4}\left(\frac{e^{-r^{2}}}{2}\sum_{n=0}^{\infty}\frac{r^{2n}}{n !}\right)^{2} = \frac{1}{16}.
\eeq

Consider the following annihilation and creation operators $\mathcal A, \mathcal A^{\dag}$  given on $\C^{2} \otimes \mathfrak H$ such that 

\beq
\mathcal A |\chi^{\pm} \otimes \Phi_{n}\rangle  = \sqrt{n \omega_{\pm}(\gamma)}|\chi^{\pm} \otimes \Phi_{n-1}\rangle, 
\quad \mathcal A^{\dag} |\chi^{\pm} \otimes \Phi_{n}\rangle  = \sqrt{(n+1) \omega_{\pm}(\gamma)}|\chi^{\pm} \otimes \Phi_{n+1}\rangle  
\eeq

and set
\beq
\mathcal F_{+}(\tilde r) = \frac{e^{\tilde r^{2}/\omega_{+}(\gamma)}}
{e^{\tilde r^{2}/\omega_{+}(\gamma)} + e^{\tilde r^{2}/\omega_{-}(\gamma)}}, \qquad 
\mathcal F_{-}(\tilde r) = \frac{e^{\tilde r^{2}/\omega_{-}(\gamma)}}
{e^{\tilde r^{2}/\omega_{+}(\gamma)} + e^{\tilde r^{2}/\omega_{-}(\gamma)}}.
\eeq

In the case of the QVCS (\ref{spoqvcs23}), we get in the quantum states $|\mathfrak Q;\pm\rangle_{1}$ the following relations 

\beq
\langle \mathcal A \rangle_{|\mathfrak Q;\pm\rangle_{1}} 
&=&  \tilde r \mathcal F_{\pm}(\tilde r)(\cos{(\vartheta)} \pm \imath \sin{(\vartheta)}\cos{(\varphi )}),
\cr 
\langle \mathcal A^{\dag} \rangle_{|\mathfrak Q;\pm\rangle_{1}} 
 &=& 
 \tilde r \mathcal F_{\pm}(\cos{(\vartheta)} \mp \imath \sin{(\vartheta)}\cos{(\varphi)}),
\eeq

\beq
\langle \mathcal Q \rangle_{|\mathfrak Q;\pm\rangle_{1}} 
&=& \tilde r \sqrt{2} \mathcal F_{\pm}(\tilde r) \cos{(\vartheta)}, \;\; \langle \mathcal P \rangle_{|\mathfrak Q;\pm\rangle_{1}}
 = \pm \tilde r \sqrt{2} \mathcal F_{\pm}(\tilde r) \sin{(\vartheta)} \cos{(\varphi)}, 
\eeq

\beq
\langle \mathcal A\mathcal A^{\dag} \rangle_{|\mathfrak Q;\pm\rangle_{1}}  = (\tilde r^{2} +  \omega_{\pm}(\gamma) ) 
\mathcal F_{\pm}(\tilde r), \quad  \langle \mathcal N \rangle_{|\mathfrak Q;\pm\rangle_{1}} = \tilde r^{2} \mathcal F_{\pm}(\tilde r), \quad 
\mathcal N = 
\mathcal A^{\dag} \mathcal A, 
\eeq

where $\mathcal Q  = \frac{1}{\sqrt{2}}(\mathcal A + \mathcal A^{\dag})$ and $\mathcal P  = \frac{1}{\imath \sqrt{2}}(\mathcal A - \mathcal A^{\dag})$.

The expectations of the operators $\mathcal Q^{2}$ and $\mathcal P^{2}$ are obtained as  
\beq
\langle \mathcal Q^{2} \rangle_{|\mathfrak Q;\pm\rangle_{1}} = 
\frac{1}{2}[4\tilde r^{2} \cos^{2}{(\vartheta)} + \omega_{\pm}(\gamma)]\mathcal F_{\pm}(\tilde r), 
\quad \langle \mathcal P^{2} \rangle_{|\mathfrak Q;\pm\rangle_{1}} = 
\frac{1}{2}[4\tilde r^{2} \sin^{2}{(\vartheta)} + \omega_{\pm}(\gamma)]\mathcal F_{\pm}(\tilde r).
\eeq

There follow the dispersions given by

\beq
(\Delta  \mathcal Q)^{2}_{|\mathfrak Q;\pm\rangle_{1}}  &=& 2 \tilde r^{2} \mathcal F_{+}(\tilde r)\mathcal F_{-}(\tilde r)  \cos^{2}{(\vartheta)} + 
\omega_{\pm}(\gamma)\frac{\mathcal F_{\pm}(\tilde r)}{2}\cr
(\Delta  \mathcal P)^{2}_{|\mathfrak Q;\pm\rangle_{1}}  &=& 
2 \tilde r^{2}\mathcal F_{\pm}(\tilde r)  \sin^{2}{(\vartheta)}\left[1-\mathcal F_{\pm}(\tilde r)\cos^{2}{(\varphi )}\right] + 
\omega_{\pm}(\gamma)\frac{\mathcal F_{\pm}(\tilde r)}{2}.
\eeq

If we take $\omega_{+}(\gamma) = \omega_{-}(\gamma)$, then $\mathcal F_{+}(\tilde r)= \frac{1}{2} = \mathcal F_{-}(\tilde r)$.  Then, 

\beq
(\Delta  \mathcal Q)^{2}_{|\mathfrak Q;\pm\rangle_{1}} &=& \frac{1}{4} [\omega_{\pm}(\gamma) + 2\tilde r^{2} \cos^{2}(\vartheta)] \cr
  (\Delta  \mathcal P)^{2}_{|\mathfrak Q;\pm\rangle_{1}} &=& \frac{1}{4} \left[\omega_{\pm}(\gamma) +
 4\tilde r^{2} \sin^{2}(\vartheta)\left(1-\frac{1}{2} \cos^{2}(\varphi )\right)\right]. 
\eeq

Assuming that 
$\tilde r^{2} \cos^{2}{(\vartheta)} \geq \frac{1}{2}$  and $\tilde r^{2} \sin^{2}{(\vartheta)} \geq \frac{1}{2}$, we get the following 
relations

\beq
(\Delta  \mathcal Q)^{2}_{|\mathfrak Q;\pm\rangle_{1}}  &\geq & \frac{1}{4} [\omega_{\pm}(\gamma) + 1] \geq \frac{1}{4} \cr
(\Delta  \mathcal P)^{2}_{|\mathfrak Q;\pm\rangle_{1}} &\geq & \frac{1}{4} [\omega_{\pm}(\gamma) + 1] \geq \frac{1}{4}
\eeq

which yield   $(\Delta \mathcal Q)^{2}_{|\mathfrak Q;\pm\rangle_{1}}(\Delta \mathcal P)^{2}_{|\mathfrak Q;\pm\rangle_{1}} \geq \frac{1}{16} $.

\subsubsection{Time evolution}

The QVCS  given in (\ref{spo12}) and  (\ref{spoqvcs23}) can be equipped  with a parameter  $\tau$  such that they are
 rewritten as

\beq{\label{spo23}}
|\mathfrak q;\pm, \tau\rangle_{1}
&=& \mathbb U(\tau)|\mathfrak q;\pm\rangle_{1} \cr
&=& \left(\mathcal N_{(\star)}(|\mathfrak q|)\right)^{-1/2} \sum_{n=0}^{\infty}\frac{\mathfrak q^{n}}{\sqrt{\rho_{(\star)}(n)}}
e^{-i  \tau \mathcal E^{\pm}_{n}} |\chi^{\pm} \otimes \Phi_{n}\rangle, 
\eeq

and 

\beq{\label{spo24}}
|\mathfrak Q;\pm,\tau\rangle_{1} =  \mathbb U(\tau)|\mathfrak Q;\pm \rangle_{1} =  (\mathcal N(\tilde r))^{-1/2}\sum_{n=0}^{\infty} (R(n))^{-1/2}
 \mathfrak Q^{n}  e^{-i  \tau \mathcal E^{\pm}_{n}}|\chi^{\pm} \otimes \Phi_{n}\rangle
\eeq

where $\mathbb U(\tau) = e^{-i  \tau \mathbb H^{red}_{{1}_{SO}}}$. 
The parameter $\tau$ is introduced such that the states  (\ref{spo23}) and (\ref{spo24}), relatively to the   time evolution operator 
$\mathbb U(t) =  e^{-i  t \mathbb H^{red}_{{1}_{SO}}}$, 
fulfill the following property

\beq
\mathbb U(t)|\mathfrak q;\pm, \tau\rangle_{1} = e^{-i  t \mathbb H^{red}_{{1}_{SO}}}|\mathfrak q;\pm, \tau\rangle_{1} = 
|\mathfrak q;\pm, \tau + t\rangle_{1},
\eeq

\beq
\mathbb U(t)|\mathfrak Q;\pm,\tau \rangle_{1} = e^{-i  t \mathbb H^{red}_{{1}_{SO}}}|\mathfrak Q;\pm,\tau \rangle_{1} 
= |\mathfrak Q;\pm,\tau+ t\rangle_{1}. 
\eeq

\section*{Conclusion}
We have first investigated,  in section $2$,  a Hamiltonian  describing an electron gas in a constant  electromagnetic field with Zeeman coupling 
interaction. A more general Hamiltonian with Rashba and Dresselhaus spin-orbit interactions has been also analysed  with a generalization 
of these both interactions, denoted by $V_{SO}$, provided. Then, in section $3$, we have  provided   the canonical  QVCS related to the studied 
  Hamiltonian   
  and then discussed the VCS  construction with their  connection with the QVCS. 
Some interesting physical features of the constructed vectors, arising from the introduction of generalized annihilation and creation  
operators which allow the definitions  on the Hilbert space $\C^{2} \otimes \mathfrak H$  of quadrature operators   
have been displayed; expectations of these operators and uncertainty relation have been computed. With the method of construction developed 
here, it is possible to extend the study in the situation of magnetoresistivity and inelastic light-scattering experiments \cite{lip-bar} where 
the interplay between Rashba, Dresselhaus, and Zeeman interactions in a quantum well submitted to an external magnetic field has been studied.

\section*{Acknowledgments}
This work is partially supported by the ICTP through the
OEA-ICMPA-Prj-15. The ICMPA is in partnership with the Daniel
Iagolnitzer Foundation (DIF), France.

\end{document}